%% file: tayler.tex
\documentclass[onecolumn]{aastex62}

\usepackage{amsmath}
\usepackage{amssymb}

\graphicspath{{./}{figures/}}

\submitjournal{ApJ}

\shorttitle{The Anelastic Tayler Instability}
\shortauthors{Goldstein et al.}

\begin{document}


\title{The Tayler Instability in the Anelastic Approximation}


\correspondingauthor{Jacqueline Goldstein}
\email{jgoldstein@astro.wisc.edu}


\author{J. Goldstein}
\affil{Department of Astronomy, University of Wisconsin-Madison,
  2535 Sterling Hall, 475 N. Charter Street, Madison, WI 53706, USA}
\affil{Kavli Institute for Theoretical Physics, University of California, Santa Barbara, CA 93106, USA}

\author{R. H. D. Townsend }
\affil{Department of Astronomy, University of Wisconsin-Madison,
  2535 Sterling Hall, 475 N. Charter Street, Madison, WI 53706, USA}
\affil{Kavli Institute for Theoretical Physics, University of California, Santa Barbara, CA 93106, USA}

\author{E. G. Zweibel}
\affil{Department of Astronomy, University of Wisconsin-Madison,
  2535 Sterling Hall, 475 N. Charter Street, Madison, WI 53706, USA}
\affiliation{Department of Physics, University of Wisconsin-Madison,
  2535 Chamberlin Hall, 1150 University Avenue, Madison, WI 53706, USA}


\begin{abstract}

The Tayler instability (TI) is a non-axisymmetric linear instability
of an axisymmetric toroidal magnetic field in magnetohydrostatic
equilibrium (MHSE). In a differentially rotating radiative region of a star, the TI could drive the Tayler-Spruit dynamo, which generates magnetic fields that can significantly impact stellar structure and evolution.
Heuristic prescriptions disagree on the efficacy of the dynamo, and numerical simulations have yet to definitively agree upon its existence. 
Criteria for the TI to develop were derived using
fully-compressible magneto-hydrodynamics, while numerical simulations
of dynamical processes in stars frequently use an anelastic
approximation. This motivates us to derive new anelastic Tayler
instability criteria. We find that some MHSE configurations
unstable in the fully-compressible case, become stable in the
anelastic case. We find and characterize the unstable modes of a
simple family of cylindrical MHSE configurations using numerical
calculations, and discuss the implications for fully non-linear
anelastic simulations.

\end{abstract}




\input{symbols.tex}


\input{introduction.tex}


\input{energy_principle.tex}


\input{compressible_vs_lbr_mhd.tex}


\input{instability_analysis.tex}


\input{instability_calcs.tex}


\input{conclusion.tex}


\acknowledgements 
\input{acknowledgements.tex}


\bibliographystyle{aasjournal}
\bibliography{tayler}


\input{appendix.tex}

\end{document}

%% file: symbols.tex

\newcommand{\gyre}{GYRE}


\newcommand{\vrad}{\boldsymbol{r}}
\newcommand{\vphi}{\boldsymbol{\phi}}

\newcommand{\vB}{\boldsymbol{B}}
\newcommand{\vJ}{\boldsymbol{J}}
\newcommand{\vg}{\boldsymbol{g}}
\newcommand{\vu}{\boldsymbol{u}}
\newcommand{\veta}{\boldsymbol{\eta}}

\newcommand{\vv}{\boldsymbol{v}}
\newcommand{\vw}{\boldsymbol{w}}

\newcommand{\vxi}{\boldsymbol{\xi}}
\newcommand{\vF}{\boldsymbol{F}}

\newcommand{\vn}{\boldsymbol{n}}

\newcommand{\xir}{\xi_{r}}
\newcommand{\xit}{\xi_{\phi}}
\newcommand{\xiz}{\xi_{z}}

\newcommand{\Br}{B_{r}}
\newcommand{\Bt}{B_{\phi}}
\newcommand{\Bz}{B_{z}}

\newcommand{\Jr}{J_{r}}
\newcommand{\Jt}{J_{\phi}}
\newcommand{\Jz}{J_{z}}

\newcommand{\txir}{\tilde{\xi}_{r}}
\newcommand{\txit}{\tilde{\xi}_{\phi}}
\newcommand{\txiz}{\tilde{\xi}_{z}}

\newcommand{\tBr}{\tilde{B}_{r}}
\newcommand{\tBt}{\tilde{B}_{\phi}}
\newcommand{\tBz}{\tilde{B}_{z}}

\newcommand{\tJr}{\tilde{J}_{r}}
\newcommand{\tJt}{\tilde{J}_{\phi}}
\newcommand{\tJz}{\tilde{J}_{z}}

\newcommand{\trho}{\tilde{\rho}}
\newcommand{\tPi}{\tilde{\Pi}}
\newcommand{\tS}{\tilde{S}}
\newcommand{\tvarpi}{\tilde{\varpi}}

\newcommand{\tr}{\tilde{r}}
\newcommand{\tra}{\tr_{\rm a}}
\newcommand{\trb}{\tr_{\rm b}}

\newcommand{\tHP}{\tilde{H}_{P}}
\newcommand{\tgr}{\tilde{g}_{r}}

\newcommand{\omegaalf}{\omega_{\rm A}}

\newcommand{\tomega}{\tilde{\omega}}
\newcommand{\tomegaalf}{\tilde{\omega}_{\rm A}}
\newcommand{\tomegamin}{\tilde{\omega}_{\rm min}}

\newcommand{\growth}{\lambda}
\newcommand{\growthmax}{\lambda_{\rm max}}

\newcommand{\mA}{\boldsymbol{\mathsf{A}}}

\newcommand{\Valf}{V_{\rm A}}


\newcommand{\diff}{\mathrm{d}}
\newcommand{\ii}{\mathrm{i}}

%% file: introduction.tex
\section{Introduction} \label{sec:intro}

The Tayler instability \citep[TI;][]{Tayler_1973, Markey_Tayler_1973, Markey_Tayler_1974, Acheson_1978, Pitts_Tayler_1985} is a
local non-axisymmetric linear instability of an axisymmetric toroidal magnetic field in magnetohydrostatic equilibrium
(MHSE). \citet{Spruit_1999} has argued that this instability is
particularly important because it can manifest when other
instabilities are suppressed by thermal stratification.  Growth rates
are on the order of the global Alfv\'{e}n-wave crossing time, which is
generally short compared to other stellar time scales, even for weak
magnetic fields. These two qualities make the TI the most relevant
magnetic instability of a toroidal magnetic field in MHSE, at least in
a non-rotating star. 

The TI has been proposed as a mechanism for significantly affecting a stars' %
structure and 
rotational evolution. \cite{Auriere_2007} proposed the TI as a mechanism for explaining the observed dichotomy in the surface magnetic field strengths of intermediate-mass stars. In stars with a relatively weak poloidal field, differential rotation generates a toroidal field unstable to the TI, transforming the field components from low-order to high-order, and yielding disk-average surface fields that may fall below observational detection thresholds. Conversely, in stars with a relatively strong poloidal field, differential rotation decays before generating a toroidal field unstable to the TI, preserving the field components, and yielding low order surface fields seen in Ap/Bp stars.


\citet{Spruit_2002} proposed the TI as one half of a dynamo that could be a significant mechanism of angular momentum transport inside the radiative regions of stars. A toroidal field unstable to the TI generates a radial field displacement which is then rewound by differential rotation back into a toroidal field, creating a dynamo loop. The magnetic torque generated by this Tayler-Spruit dynamo could be a missing link in stellar evolution theory, where there is currently a discrepancy between the modeled and observed rotation rates of red giant cores, and of stellar remnants.

\citet{Cantiello_etal_2014} showed that the heuristic prescription for
the Tayler-Spruit dynamo implemented in the stellar evolution code
Modules for Experiments in Stellar Astrophysics \citep[MESA;][]{Paxton_etal_2013,Paxton_etal_2015,Paxton_etal_2018} increases angular momentum transport during red giant branch evolution. The results cannot fully
explain the slow core rotation rates of red giant branch stars as observed by \textit{Kepler}, but the models with the dynamo are in better agreement with observations than the models without.  Recently \cite{Fuller_2019} demonstrated that a revised prescription implemented into MESA could largely reproduce observed rotation rates.

\citet{Maeder_2003, Maeder_2004, Maeder_2005} showed that a heuristic prescription for the Tayler-Spruit dynamo, as implemented in the Geneva stellar evolution code \citep{Meynet_2005}, can have a significant effect on the main-sequence evolution of massive stars. The dynamo imposes near solid-body rotation that enhances meridional circulation and efficient mixing, resulting in larger convective cores, longer main-sequence life times, enriched surface abundances of nucleosynthesized elements, and elevated stellar luminosities. \citet{Song_2016, Song_2018} demonstrated that for massive stars in binary systems, spun up through tidal interactions, dynamo-induced solid body rotation can lead to 
similar outcomes.

Despite the potential significance of the Tayler-Spruit dynamo in stellar structure and evolution, the existence and nature of the dynamo is
currently debated through both analytical and numerical
calculations. Analytically, various heuristic prescriptions have been
developed to predict the magnitude of the magnetic torque
\citep{Spruit_1999, Spruit_2002, Maeder_2003, Maeder_2004, Heger_etal_2005, Braithwaite_2006a, Denissenkov_Pinsonneault_2007}. Numerically,
non-linear MHD simulations have been unable to agree whether the dynamo actually operates as envisaged \citep{Braithwaite_2006,
  Zahn_etal_2007}.
  
In light of the potential importance of the Tayler-Spruit dynamo for
stars, the disagreement between the analytical predictions
and the numerical results is unsettling.  
The discrepancy motivates us to look at the basic assumptions used in the original TI critera and in the non-linear MHD simulations investigating the Tayler-Spruit dynamo. \cite{Tayler_1973} developed criteria for the TI using fully-compressible ideal MHD, while 
numerical MHD simulations of stellar interiors frequently use anelastic MHD (eg. \cite{Brown_etal_2012}), an approximation that
filters out sound waves, which are very short-period relative to stellar
timescales and are therefore prohibitively expensive to compute.

The goal of this paper is to re-examine the TI in the anelastic
approximation. We derive new anelastic TI (anTI) stability criteria and apply them to a family of simple MHSE models to
determine which are subject to the instability. We verify our results
numerically using a modified version of the \gyre\ stellar oscillation
code \citep{Townsend_Teitler_2013}, which solves a system of
linearized anelastic MHD equations to calculate growth rates and
eigenfunctions of unstable modes. We conclude that the anelastic case
is more restrictive, but that the TI should be present in anelastic
MHD simulations if the models used are unstable under the anTI
criteria.

The paper is structured as follows. In
Section~\ref{sec:energy_principle} we given an overview of the energy
principle --- the method used to develop the instability criteria. In
Section~\ref{sec:compressible_vs_lbr_mhd} we introduce the
fully-compressible MHD equations and the Lantz-Braginsky-Roberts (LBR)
anelastic approximation for MHD, a form that is valid in the
isothermal atmosphere we assume in our later analysis.
In Section~\ref{sec:instability_analysis} we summarize the original
  TI stability criteria derived from the energy principle, and derive
  the new anTI criteria. In Section~\ref{sec:instability_calcs} we compare
  the original and anTI criteria to \gyre's numerically calculated
  growth rates and eigenfunctions for unstable modes in our models,
  showing that the anTI criteria are correct in anelastic MHD. In
  Section~\ref{sec:conclusion} we conclude with considerations for future
  analytical and numerical work.

%% file: energy_principle.tex
\section{Energy Principle}
\label{sec:energy_principle}

The instability analysis in \cite{Tayler_1973} was developed using the
MHD energy principle of \cite{Bernstein_etal_1958}, which gives the
necessary and sufficient condition for an energy-conserving, ideal
system in MHSE to be unstable to small displacements. Although the
energy principle is widely used in studies of laboratory and natural
plasmas, we will need to modify it to accommodate the anelastic
equations, so we briefly review it here.

The energy principle is based on being able to write the linearized
equation of motion for the fluid displacement perturbation, $\vxi$, in
the form
\begin{equation}\label{eq:motion}
\rho\frac{\partial^2 \vxi}{\partial t^2}= \vF(\vxi),
\end{equation}
where $\vxi$ is related to velocity perturbations $\vu'$ by
\begin{gather}\label{eq:xi}
\vu' = \frac{\partial \vxi}{\partial t},
\end{gather}
and $\vF$ is a linear, self adjoint operator\footnote{By self
  adjointness we mean $\int \vxi \cdot \vF(\veta) \,\diff\tau = \int
  \veta \cdot \vF(\vxi) \,\diff\tau$ for displacement vectors $\vxi$,
  $\veta$ that obey suitable boundary conditions.}. Since
equation~(\ref{eq:motion}) doesn't explicitly depend on time, we look
for separable solutions of the form
\begin{gather} \label{eq:xi_time}
  \vxi(t) \propto \exp{\ii \omega t},
\end{gather}
where $\omega$ is an angular frequency. It follows from the self
adjointness property that $\omega^2$ is real; $\omega^2 < 0$ signifies
instability, with an exponential growth rate $\growth = |\omega|$.

It can be shown that equation~(\ref{eq:motion}), together with the self
adjointness property, leads to a conservation law which we identify
with conservation of perturbation energy:
\begin{equation}\label{eq:conservedE}
  \frac{\partial}{\partial t} \left[
    \frac{1}{2} \int \rho \frac{\partial \vxi}{\partial t} \cdot \frac{\partial\vxi}{\partial t} \,\diff\tau -
    \frac{1}{2} \int \vxi \cdot \vF(\vxi) \,\diff\tau \right]=0.
\end{equation}
The first term in brackets in equation~(\ref{eq:conservedE}) is the
kinetic energy $\delta K$, while the second represents the
potential energy
\begin{gather} \label{eq:PE}
\delta W =  -\frac{1}{2} \int \vxi \cdot \vF(\vxi) \,\diff\tau.
\end{gather}
Equation~(\ref{eq:conservedE}) shows that $\delta K + \delta W$ is
constant in time.  For an unstable mode, both $\delta K$ and
$\delta W$ grow exponentially in magnitude with time, and $\delta K$
is positive definite; therefore, $\delta W$ must be negative. This is the basis of the energy principle.

The energy principle has both advantages and disadvantages compared to
solving for the eigenvalues $\omega^{2}$ of
equation~(\ref{eq:motion}). The advantage is that it is often easier
to minimize $\delta W$, or even evaluate it for a set of trial
functions, than to solve the coupled set of differential equations
corresponding to the eigenvalue problem and hope to capture all
modes. The disadvantage is that the energy principle yields at
best the lowest-eigenvalue  mode (obtained by rigorous minimization of
$\delta W$) rather than the whole spectrum of stable and unstable
modes.

%% file: compressible_vs_lbr_mhd.tex
\section{Fully Compressible and Anelastic Ideal MHD} \label{sec:compressible_vs_lbr_mhd}

Fully compressible, isentropic, ideal MHD is described by the
conservation equations for mass, momentum, and entropy, together with Faraday-Maxwell's Law
(combined with Ohm's Law), Amp\`ere's Law, and Gauss' Law for
magnetism,
\begin{gather}
\label{eq:continuity}
\frac{\partial \rho}{\partial t} + \nabla \cdot (\rho \vu) = 0, \\
\label{eq:momentum}
\frac{\partial \vu}{\partial t} + \vu \cdot \nabla \vu =
-\frac{1}{\rho}\nabla P + \vg +  \frac{\vJ \times \vB}{\rho c}, \\
\label{eq:entropy}
\frac{\partial S}{\partial t} + \vu \cdot \nabla S = 0, \\  
\label{eq:faraday}
\frac{\partial \vB}{\partial t} = \nabla \times \left( \vu \times \vB \right), \\
\label{eq:ampere}
\vJ = \frac{c}{4\pi} \nabla \times \vB, \\
\label{eq:gauss}
\nabla \cdot \vB = 0.
\end{gather}
Here, $\vu$, $\rho$, $P$, $S$, $\vg$, $\vB$ and $\vJ$ are the fluid
velocity, density, pressure, specific entropy, gravitational
acceleration, magnetic field and current density, respectively. The
pressure, density and specific entropy are assumed to be related by
the equation of state
\begin{equation}  \label{eq:state}
\rho = \rho(S, P),
\end{equation}
which for simplicity we assume to follow ideal-gas behavior.

In order to model the evolution of small perturbations of a static
equilibrium state, we decompose the dependent variables into the sum of
a background value (denoted by the subscript 0), and a perturbed value
(denoted by a prime):
\begin{equation}
  \label{eq:lin_decomp}
\begin{aligned}
\vu &= \vu', \\
P &= P_0 + P', \\
\rho &= \rho_0+ \rho',\\
S &= S_0 + S', \\
\vB &= \vB_0 + \vB', \\
\vJ &= \vJ_0 + \vJ'
\end{aligned}
\end{equation}
we neglect any perturbations to the gravity $\vg$). The background values obey the MHSE condition
\begin{equation} \label{eq:mhse}
  \frac{\nabla P_0}{\rho_0} = \vg + \frac{\vJ_0 \times \vB_0}{\rho_0 c}, 
\end{equation}
the equilibrium current equation
\begin{equation}
  \vJ_0 = \frac{c}{4\pi} \nabla \times \vB_0,
\end{equation}
and a gradient relation that follows from the equation of state~(\ref{eq:state}),
\begin{equation} \label{eq:gradients}
  \frac{\nabla \rho_{0}}{\rho_{0}} = \frac{\nabla P_{0}}{\gamma P_{0}} - \frac{\nabla S_{0}}{c_P},
\end{equation}
where $c_{P}$ is the specific heat at constant pressure and $\gamma$
is the ratio of specific heats. Likewise, the perturbed values
satisfy the linearized ideal MHD equations that follow when we
substitute the expressions~(\ref{eq:lin_decomp}) into
equations~(\ref{eq:continuity}--\ref{eq:gauss}), subtract the
background state, and discard terms that are second- or higher-order in perturbed quantities:
\begin{gather}
\label{eq:lin_continuity}
\frac{\partial \rho'}{\partial t} + \nabla \cdot (\rho_{0} \vu') = 0, \\
\label{eq:lin_momentum}
\frac{\partial \vu'}{\partial t} = - \frac{\nabla P'}{\rho_{0}} + \frac{\rho'}{\rho_{0}} \vg +
\left(\frac{ \vJ' \times \vB_{0} + \vJ_{0} \times \vB'}{\rho_{0} c} \right) ,\\
\label{eq:lin_entropy}
\frac{\partial S'}{\partial t} + \vu' \cdot \nabla  S_{0} = 0,\\
\label{eq:lin_faraday}
\frac{\partial \vB'}{\partial t}= \nabla \times \left( \vu' \times \vB_{0} \right) , \\
\vJ' = \frac{c}{4 \pi }\left(\nabla \times \vB'\right), \label{eq:lin_ampere}\\
\nabla \cdot \vB' = 0. \label{eq:lin_gauss}
\end{gather}

Numerical MHD simulations that are relevant on stellar scales of
interest frequently use an anelastic approximation,
\begin{equation}
  \nabla \cdot (\rho_0 \vu') = 0,
  \label{eq:anelastic}
\end{equation}
that filters out fast, high-frequency sound waves unimportant on
stellar scales, while keeping slower internal gravity waves. The anelastic
approximation is technically valid only for adiabatically stratified
systems ($\nabla S_{0} = 0$), but is now used to study problems such as penetrative
convection and interface dynamos that include stably stratified
regions. In such contexts, \citet{Brown_etal_2012} studied energy
conservation in three widely used forms of the anelastic
equations. They showed that one, the so-called
Lantz-Braginsky-Roberts (LBR) formulation \citep{Lantz_1992,Braginsky_Roberts_1995}
conserves energy, while the others conserve a related but distinct
pseudo-energy. Therefore, we consider the LBR formulation to be the
best candidate for analyzing the Taylor instability in the anelastic
approximation, and study only that version in this paper. However, as
we will see, the energy principle has to be modified even for the LBR
formulation.

The LBR formulation re-writes the linearized momentum equation in terms of entropy and a reduced pressure perturbation
\begin{gather}
 \varpi' \equiv \frac{P'}{\rho_0},
\end{gather}
and neglects a term $\varpi' \nabla (S_{0}/c_{P})$.
Although \citet{Brown_etal_2012}
neglected the effect of magnetic fields, it can be shown by
recapitulating their analysis that neglecting \emph{all} terms
proportional to $\varpi'$ (but not its gradient) in the linearized
momentum equation preserves energy conservation in an isothermal
atmosphere, even in the presence of magnetic fields.

We develop an LBR version of the linearized
momentum equation~(\ref{eq:lin_momentum}) by eliminating the density
perturbation using the linearized equation of state,
\begin{equation}
\frac{\rho'}{\rho_{0}} = \frac{1}{\gamma} \frac{P'}{P_{0}} - \frac{S'}{c_P},
\label{eq:lin_eos}
\end{equation}
and likewise eliminating the background pressure gradient $\nabla P_{0}$
using the MHSE condition~(\ref{eq:mhse}). The linearized momentum
equation then becomes
\begin{equation}
  \frac{\partial \vu' }{\partial t} = -\nabla{\varpi'} - \frac{S'}{c_P} \vg +
  \left( \frac{\vJ' \times \vB_0 + \vJ_0 \times \vB'}{\rho_0 c} \right),
\label{eq:lbr_momentum}
\end{equation}
where, as discussed above, we have dropped all terms proportional to
$\varpi'$. The linearized LBR anelastic MHD equations then comprise equations~(\ref{eq:lin_entropy}--\ref{eq:anelastic}) and~(\ref{eq:lbr_momentum}).

%% file: instability_analysis.tex
\section{Instability Analysis} \label{sec:instability_analysis}

\subsection{Fully-Compressible Analysis} \label{sec:instability_analysis_tay}

\cite{Tayler_1973} derived his criteria for stability by
  applying a fully-compressible MHD form of the energy principle (see
  Section~\ref{sec:energy_principle}) to an equilibrium configuration
  comprising an axisymmetric toroidal magnetic field in a
  non-rotating stratified plasma. In this configuration, all
  background quantities are functions only of the cylindrical radial
  ($r$) and axial ($z$) coordinates. Here, we briefly recapitulate
his analysis. First, we integrate equations~(\ref{eq:lin_continuity})
and~(\ref{eq:lin_faraday}) with respect to time to obtain explicit
expressions for the density, magnetic field and current perturbations
in terms of the displacement vector $\vxi$,
\begin{gather}
\rho' = - \nabla \cdot (\rho_{0} \vxi), \label{eq:rho'}\\
\vB' = \nabla \times (\vxi \times \vB_{0}), \label{eq:B'}\\
\vJ' = \frac{c}{4 \pi} \nabla \times [\nabla \times (\vxi \times \vB_{0})] \label{eq:J'}.
\end{gather}
The linearized equation of state~(\ref{eq:lin_eos}) leads to a
corresponding expression for the pressure perturbation,
\begin{equation}
P' = - \gamma P_{0} \nabla \cdot \vxi - \vxi \cdot \nabla P_{0} \label{eq:p'}.
\end{equation}
By substituting these expressions into the linearized momentum
equation~(\ref{eq:lin_momentum}), the force operator introduced in
equation~(\ref{eq:motion}) is found as
\begin{equation}\label{eq:force_op_tay}
\vF(\vxi) = \nabla (\gamma P_{0} \nabla \cdot \vxi + \vxi \cdot \nabla P_{0}) - \nabla \cdot (\rho_{0} \vxi) \, \vg
+ \frac{1}{4 \pi} \left[ (\nabla \times \vB') \times \vB_{0} + (\nabla \times \vB_{0}) \times \vB' \right].
\end{equation}
\citet{Kulsrud_1964} demonstrated that this force operator is
self-adjoint under boundary conditions
\begin{equation}
\label{eq:boundary}
\vxi \cdot \hat{\vn} = 0, \qquad \vB_{0} \cdot \hat{\vn} = 0,
\end{equation}
that correspond to rigid, perfectly-conducting walls; here,
$\hat{\vn}$ is the unit surface normal vector of the boundary.

Taking the scalar product of equation~(\ref{eq:force_op_tay}) with
$\vxi$, and integrating over volume, leads via equation~(\ref{eq:PE})
to the potential energy 
\begin{equation} \label{eq:PE_tay}
  \delta W = \frac{1}{2} \int \diff\tau \left[
    \frac{\vB'^2}{4\pi} - \frac{\vJ_{0} \cdot (\vB' \times \vxi)}{c} + \gamma P_{0} (\nabla \cdot \vxi)^2 + 
    (\vxi \cdot \nabla P_{0})\ \nabla \cdot \vxi + 
    (\vxi \cdot \vg)\ \nabla \cdot (\rho_{0} \vxi) \right],
\end{equation}
where we have made use of the boundary conditions~(\ref{eq:boundary})
to eliminate surface integrals involving $\vxi \cdot \hat{\vn}$ and
$\vB_{0} \cdot \hat{\vn}$.

Following \citet{Tayler_1973}, we write the fluid displacement vector
in cylindrical coordinates as
\begin{equation}
\label{eq:xi_form}
\begin{bmatrix}
\xi_{r}      \\
\xi_{\phi} \\
\xi_{z}
\end{bmatrix}
=
\begin{bmatrix}
X \cos{m\phi} \\
-(Y/m) \sin m\phi \\
Z \cos m\phi
\end{bmatrix},
\end{equation}
where $X, Y, Z$ are real functions of $r$ and $z$, and the integer $m$
is an azimuthal wavenumber. With these definitions, we evaluate
  the $\phi$ part of the integral in equation~(\ref{eq:PE_tay}) to
  obtain
\begin{multline}
 \label{eq:PE_tay_int}  
 \delta W = \frac{1}{8} \int r \,\diff r \, \diff z
 \left\{
 \frac{m^{2} B_{\phi}^{2}}{r^{2}} \left( X^{2} + Z^{2} \right) +
 \left[ \frac{\partial}{\partial r} (B_{\phi} X) + \frac{\partial}{\partial z} (B_{\phi} Z) \right]^{2} -
 \right.
 \\
  \left[ \frac{\partial}{\partial r} (B_{\phi} X) + \frac{\partial}{\partial z} (B_{\phi} Z) \right]
  \left[ \frac{1}{r} \frac{\partial}{\partial r} (r B_{\phi}) X + \frac{\partial B_{\phi}}{\partial z} Z \right] 
  - \frac{B_{\phi}}{r} Y \left[ \frac{1}{r} \frac{\partial}{\partial r} (r B_{\phi}) X + \frac{\partial B_{\phi}}{\partial z} Z \right] +
 \\
 4\pi \gamma P_{0} \left[ \frac{1}{r} \frac{\partial}{\partial r} (r X) - \frac{Y}{r} + \frac{\partial Z}{\partial z} \right]^{2} +
 4\pi (g_{r} X + g_{z} Z ) \left( \frac{\partial \rho_{0}}{\partial r} X + \frac{\partial \rho_{0}}{\partial z} Z \right) +
 \\
 \left.
 4\pi \left[ X \left( \frac{\partial P_{0}}{\partial r} + \rho_{0} g_{r} \right) + Z \left( \frac{\partial P_{0}}{\partial z} + \rho_{0} g_{z} \right) \right]
 \left[ \frac{1}{r} \frac{\partial}{\partial r} (r X) - \frac{Y}{r} + \frac{\partial Z}{\partial z} \right]
 \right\},
\end{multline}
where $g_{r}$ and $g_{z}$ are the radial and axial components of the
background gravity $\vg$, respectively, and $B_{\phi}$ is the
azimuthal component of the background field $\vB_{0}$. This expression
is minimized with respect to $Y$ by solving 
\begin{equation}
  \frac{\partial \delta W}{\partial Y} = 0,
\end{equation}
to obtain
\begin{equation}
\label{eq:y_r_tay}
\frac{Y}{r}= \frac{1}{r} \frac{\partial}{\partial r} \left( r X \right) + \frac{\partial Z}{\partial z} + \frac{\rho_{0}}{\gamma P_{0}} \left( g_{r} X + g_{z} Z \right).
\end{equation}
Substituting this back into equation~(\ref{eq:PE_tay_int}), the
minimized potential energy is found as
\begin{equation}
  \label{eq:PE_tay_min}
  \delta W_{\rm min} = \frac{1}{8} \int  r \,\diff r \, \diff z
  \left\{
  B_{\phi}^{2} \left[ r \frac{\diff}{\diff r} \left( \frac{X}{r} \right) + \frac{\diff Z}{\diff z} \right]^{2} +
  4\pi \left[ \mathcal{A} X^{2} + \mathcal{B} X Z + \mathcal{C} Z^{2} \right]
  \right\}
\end{equation}
where we introduce
\begin{equation}
  \label{eq:coeff_tay}
  \begin{aligned}
    \mathcal{A} &\equiv -
    \rho_{0} g_{r} \left( \frac{\rho_{0} g_{r}}{\gamma P_{0}} - \frac{1}{\rho_{0}} \frac{\partial \rho_{0}}{\partial r} \right) + 
    m^{2} \frac{B_{\phi}^{2}}{4 \pi r^2} - \frac{B_{\phi}}{2\pi r^{2}}\frac{\partial}{\partial r} (r B_{\phi}), \\
    \mathcal{B} &\equiv -
    \rho_{0} g_{r} \left( \frac{\rho_{0} g_{z}}{\gamma P_{0}} - \frac{1}{\rho_{0}} \frac{\partial \rho_{0}}{\partial z} \right) -
    \rho_{0} g_{z} \left( \frac{\rho_{0} g_{r}}{\gamma P_{0}} - \frac{1}{\rho_{0}} \frac{\partial \rho_{0}}{\partial r} \right) - 
    \frac{B_{\phi}}{2 \pi r}\frac{\partial B_{\phi}}{\partial z}, \\
    \mathcal{C} &\equiv -
    \rho_{0} g_{z} \left( \frac{\rho_{0} g_{z}}{\gamma P_{0}} - \frac{1}{\rho_{0}} \frac{\partial \rho_{0}}{\partial z} \right) +
    m^{2} \frac{B_{\phi}^{2}}{4 \pi r^{2}}.
  \end{aligned}
\end{equation}
The first term in the integrand of equation~(\ref{eq:PE_tay_min}) is
positive-definite. For the remaining terms, which appear as a quadratic
form in $X$ and $Z$, the sufficient conditions that $\delta W_{\min} >
0$ are that
\begin{equation}
  \label{eq:stab_tay}
  \mathcal{A} > 0, \qquad \mathcal{C} > 0, \qquad 4 \mathcal{A} \mathcal{C} > \mathcal{B}^{2}
\end{equation}
everywhere. These are the criteria for \emph{stability} against the
TI. Since the azimuthal order $m$ appears only in the
positive-definite terms $m^{2} B_{\phi}^{2}/4 \pi r^{2}$ in equation~(\ref{eq:coeff_tay}), the most
unstable non-axisymmetric modes correspond to $|m| = 1$. For these
modes, the above expressions reduce to the criteria given by
\citet[][his equations~2.20--2.22]{Tayler_1973}, modulo factors of
$4\pi$ that arise from our choice of electromagnetic units. As Tayler
demonstrates in his Appendix 2, the above criteria are not only sufficient
but also necessary, in that violation of one or more of these
inequalities can lead to instability ($\delta W_{\rm min} < 0$) for a
suitable choice of $X$ and $Z$.

\subsection{Constrained Analysis} \label{sec:instability_analysis_const}

We now consider how to implement the anelastic constraint in our
analysis. This constraint removes the freedom to choose a $Y$ that
minimizes $\delta W$; instead, we must set
\begin{equation}
\label{eq:y_r_const}
\frac{Y}{r} = \frac{1}{r}\frac{\partial}{\partial r}(rX) + \frac{\partial Z}{\partial z} +
\frac{1}{\rho_{0}} \frac{\diff \rho_{0}}{\diff r} X +
\frac{1}{\rho_{0}} \frac{\diff \rho_{0}}{\diff z} Z
\end{equation}

to ensure that equation~(\ref{eq:anelastic}) is satisfied. We repeat
the analysis of the preceding section, but using this expression in
place of equation~(\ref{eq:y_r_tay}); this yields a potential energy
that is identical to equation~(\ref{eq:PE_tay_min}), save that the
quadratic-form coefficients $\mathcal{A}$, $\mathcal{B}$ and
$\mathcal{C}$ are replaced by
\begin{equation}
  \label{eq:coeff_c}
  \begin{aligned}
  \mathcal{A}_{\rm c} &\equiv \mathcal{A} + \gamma P_{0}
  \left( \frac{\rho_{0} g_{r}}{\gamma P_{0}} - \frac{1}{\rho_{0}} \frac{\partial \rho_{0}}{\partial r} \right)^{2}, \\
  \mathcal{B}_{\rm c} &\equiv \mathcal{B} + 2 \gamma P_{0}
  \left( \frac{\rho_{0} g_{r}}{\gamma P_{0}} - \frac{1}{\rho_{0}} \frac{\partial \rho_{0}}{\partial r} \right)
  \left( \frac{\rho_{0} g_{z}}{\gamma P_{0}} - \frac{1}{\rho_{0}} \frac{\partial \rho_{0}}{\partial z} \right), \\
  \mathcal{C}_{\rm c} &\equiv \mathcal{C} + \gamma P_{0}
  \left( \frac{\rho_{0} g_{z}}{\gamma P_{0}} - \frac{1}{\rho_{0}} \frac{\partial \rho_{0}}{\partial z} \right)^{2},
  \end{aligned}
\end{equation}
respectively (here, the `c' subscripts stand for `constrained'). The
criteria for stability are now that
\begin{equation}
  \label{eq:stab_c}
  \mathcal{A}_{\rm c} > 0, \qquad \mathcal{C}_{\rm c} > 0, \qquad 4 \mathcal{A}_{\rm c} \mathcal{C}_{\rm c} > \mathcal{B}_{\rm c}^{2}.
\end{equation}
As we shall demonstrate in Section~\ref{sec:instability_calcs}, these
constrained TI (cTI) criteria under-predict the extent of the
instability found by numerical calculations employing the anelastic
condition. This shortcoming motivates a more careful treatment, based
on re-deriving the force operator from the LBR anelastic linearized
momentum equation~(\ref{eq:lbr_momentum}).

\subsection{LBR Anelastic Analysis} \label{sec:instability_analysis_lbr}

To re-derive the force operator in the LBR anelastic case,
  first we integrate equation~(\ref{eq:lin_entropy}) with respect to
  time to obtain
\begin{gather}
S' = - \vxi \cdot \nabla S_{0}.
\end{gather}
Substituting this expression and
equations~(\ref{eq:rho'}--\ref{eq:p'}) into the linearized momentum
equation~(\ref{eq:lbr_momentum}), the LBR anelastic force operator is
derived as
\begin{equation} \label{eq:force_op_lbr}
\vF_{\rm LBR}(\vxi) = 
- \rho_{0} \nabla \varpi'  - \rho_{0} \vg \, \vxi \cdot \left[ \frac{\nabla P_{0}}{\gamma P_{0}} -
  \frac{\nabla \rho_{0}}{\rho_{0}} \right] + 
\frac{1}{4 \pi} \left[ (\nabla \times \vB' ) \times \vB_{0} + (\nabla \times \vB_{0}) \times \vB' \right],
\end{equation}
where we have used equation~(\ref{eq:gradients}) to eliminate the
background entropy gradient $\nabla S_{0}$. This operator is
self-adjoint under the same boundary conditions~(\ref{eq:boundary}) as
applied before.

We repeat the analysis of
Section~\ref{sec:instability_analysis_tay}, but using
equations~(\ref{eq:y_r_const},\ref{eq:force_op_lbr}) in place
of~(\ref{eq:force_op_tay},\ref{eq:y_r_tay}). The resulting
  potential energy is identical to equation~(\ref{eq:PE_tay_min}),
  save that the quadratic-form coefficients $\mathcal{A}$,
  $\mathcal{B}$ and $\mathcal{C}$ are now replaced by
\begin{equation}
  \label{eq:coeff_lbr}
  \begin{aligned}
  \mathcal{A}_{\rm LBR} &\equiv \mathcal{A} +
  \frac{B_{\phi}}{4\pi r} \frac{\partial}{\partial r} (r B_{\phi}) 
  \left( \frac{\rho_{0} g_{r}}{\gamma P_{0}} - \frac{1}{\rho_{0}} \frac{\partial \rho_{0}}{\partial r} \right), \\
  \mathcal{B}_{\rm LBR} &\equiv \mathcal{B} +
  \frac{B_{\phi}}{4\pi} \frac{\partial B_{\phi}}{\partial z}
  \left( \frac{\rho_{0} g_{r}}{\gamma P_{0}} - \frac{1}{\rho_{0}} \frac{\partial \rho_{0}}{\partial r} \right) +
  \frac{B_{\phi}}{4\pi r} \frac{\partial}{\partial r} (r B_{\phi}) 
  \left( \frac{\rho_{0} g_{z}}{\gamma P_{0}} - \frac{1}{\rho_{0}} \frac{\partial \rho_{0}}{\partial z} \right), \\
  \mathcal{C}_{\rm LBR} &\equiv \mathcal{C} +
  \frac{B_{\phi}}{4\pi} \frac{\partial B_{\phi}}{\partial z}
  \left( \frac{\rho_{0} g_{z}}{\gamma P_{0}} - \frac{1}{\rho_{0}} \frac{\partial \rho_{0}}{\partial z} \right),
  \end{aligned}
\end{equation}
respectively. The criteria for stability against the TI are now that
\begin{equation}
  \label{eq:stab_lbr}
  \mathcal{A}_{\rm LBR} > 0, \qquad \mathcal{C}_{\rm LBR} > 0, \qquad 4 \mathcal{A}_{\rm LBR} \mathcal{C}_{\rm LBR} > \mathcal{B}_{\rm LBR}^{2}.
\end{equation}
Using the same approach as in Appendix 2 of \citet{Tayler_1973}, we
can show that these anelastic TI (anTI) criteria are necessary as well
as sufficient. We note that the most unstable non-axisymmetric modes
still correspond to $|m|=1$. Together, equations~(\ref{eq:coeff_lbr})
and~(\ref{eq:stab_lbr}) make up the principal result of this paper.

%% file: instability_calcs.tex
\section{Numerical Instability Calculations} \label{sec:instability_calcs}

In this section, we compare the analytic work in the preceding
sections against numerical solutions of the linearized LBR anelastic
MHD equations and boundary conditions. Details of our numerical
technique are given in the Appendix; in brief, we modify the
\gyre\ stellar oscillation code \citep{Townsend_Teitler_2013} to find
the modal eigenvalues and eigenfunctions. Because the
  \gyre\ code is restricted to solving 1D eigenproblems, we focus our
  analysis on a reduced equilibrium case in which background
  quantities depend only on $r$.

\subsection{\protect Equilibrium Model} \label{ssec:instability_calcs_model}

The 1D equilibrium model we consider assumes an isothermal
stratification characterized by a constant sound speed $a$ and a
constant ratio $\beta$ of gas pressure to magnetic pressure. Radial
gravity is provided by a line mass on the cylindrical axis of
symmetry,
\begin{equation}
  g_{r}(r) = - \frac{q a^2}{r} \label{eq:gr}.
\end{equation}
where $q \geq 0$ is a dimensionless gravitational strength parameter.
Solving the MHSE equation~(\ref{eq:mhse}) yields a power-law
background pressure distribution
\begin{equation}
P_{0}(r) = P_{0,0} \left(\frac{r}{r_{0}}\right)^{-\alpha},
\label{eq:background-P}
\end{equation}
where $P_{0,0}$ is the pressure at some fiducial radius $r_{0}$, and
\begin{equation}
\alpha \equiv \frac{2 + q \beta}{ 1 + \beta}.
\label{eq:alpha}
\end{equation}
The background density and azimuthal field strength are
\begin{equation}
  \rho_{0}(r) = \frac{P_{0}(r)}{a^{2}}
\end{equation}
and
\begin{equation}
  B_{\phi}(r) = \left[ \frac{8 \pi P_{0}(r)}{\beta} \right]^{1/2},
\end{equation}
respectively.

It can be shown that for $\alpha < 2$, the magnetic tension in this
equilibrium model dominates the magnetic pressure, meaning that the
stratification is both magnetically and gravitationally confined.  To
satisfy $\alpha < 2$, from equation~(\ref{eq:alpha}), the
gravitational parameter must be $q < 1$. For stars, of course, we
expect that the magnetic field is relatively weak and that the
equilibrium is close to hydrostatic, whether the magnetic field
provides pressure support through its negative gradient or confinement
through tension.

In the context of the 1D equilibrium model described here, the
quadratic coefficients~(\ref{eq:coeff_tay}) for the stability criteria
in the fully compressible case become
\begin{equation}
  \label{eq:coeff_tay_1d}
  \mathcal{A} = \left[
    q \left( \alpha - \frac{q}{\gamma} \right) + \frac{2}{\beta} (\alpha + m^{2} - 2)
    \right] \frac{P_{0}}{r^{2}},  \qquad
  \mathcal{B} = 0, \qquad
  \mathcal{C} = \frac{2 m^{2}}{\beta} \frac{P_{0}}{r^{2}}.
\end{equation}
The stability criteria~(\ref{eq:stab_tay}) then reduce to the
requirement that the bracketed term in the expression for
$\mathcal{A}$ be positive.  

For the constrained anelastic and LBR anelastic cases, the
corresponding expressions are
\begin{equation}
  \label{eq:coeff_c_1d}
  \mathcal{A}_{\rm c} = \left[
    q \left( \alpha - \frac{q}{\gamma} \right) + \frac{2}{\beta} (\alpha + m^{2} - 2) + \frac{(q - \alpha \gamma)^{2}}{\gamma}
    \right] \frac{P_{0}}{r^{2}}, \qquad
  \mathcal{B}_{\rm c} = 0, \qquad
  \mathcal{C}_{\rm c} = \frac{2 m^{2}}{\beta} \frac{P_{0}}{r^{2}},
\end{equation}
and
\begin{equation}
  \label{eq:coeff_lbr_1d}
  \mathcal{A}_{\rm LBR} = \left[
    q \left( \alpha - \frac{q}{\gamma} \right) + \frac{2}{\beta} (\alpha + m^{2} - 2) + \frac{(\alpha - 2)(q - \alpha\gamma)}{\beta \gamma}
    \right] \frac{P_{0}}{r^{2}},  \qquad
  \mathcal{B}_{\rm LBR} = 0, \qquad
  \mathcal{C}_{\rm LBR} = \frac{2 m^{2}}{\beta} \frac{P_{0}}{r^{2}},
\end{equation}
respectively. As before, the stability criteria reduce to the
requirement that the bracketed terms are positive.

\subsection{\protect Stable and Unstable Modes} \label{ssec:instability_calcs_modes}

As an initial demonstration of our numerical solution technique, we
calculate eigenvalues and eigenfunctions of stable and unstable modes
for an equilibrium configuration having $\beta=5$, $q=0.01$ and
$\gamma=5/3$; this choice of parameters ensures that we are looking at
a robust instability, as we shall later show in a parameter study. We
solve the linearized equations and boundary conditions on a spatial
grid of 1,000 points, uniformly spanning the domain $[\tra,\trb] = [1,
  1.5]$ (here, $\tr$ is the dimensionless radius introduced in the
Appendix). Assuming an azimuthal wavenumber $m=1$ and an axial
wavenumber $k=25$, we search for modes having eigenvalues
$\tomega^{2}$ below the square $\tomegaalf^{2}$ of the dimensionless
Alfv\'en frequency evaluated at $\tr=\trb$ (see
equation~\ref{eq:dim-alf-freq}). These modes comprise an infinite
family, in which each mode can uniquely be classified by a radial
order $n$ that counts the number of nodes (excluding the
endpoints) in the dimensionless radial displacement eigenfunction
$\txir$. With this classification, an ordering by $n$ is in one-to-one
correspondence with an ordering by $\tomega^{2}$, with the eigenvalue
$\tomega^{2}_{0}$ of the $n=0$ (fundamental) mode being the least
positive.

\begin{figure}
  \centering
  \includegraphics{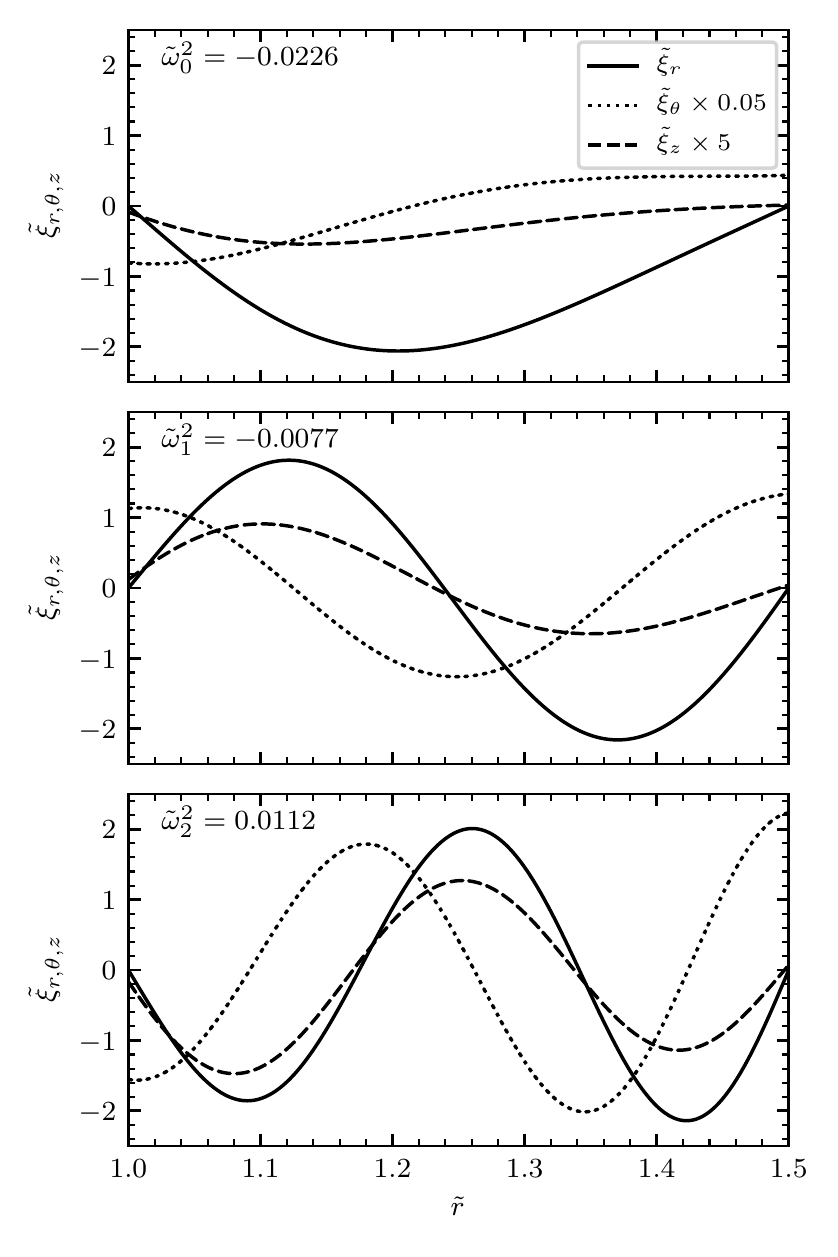}
  \caption{Dimensionless displacement eigenfunctions of the $n=0,1,2$
    modes (top to bottom), for an azimuthal wavenumber $m=1$ and an
    axial wavenumber $k=25$, and an equilibrium model having
    $\beta=5$, $q=0.01$ and $\gamma=5/3$. In each panel, the
    eigenfunctions are normalized such that the root-mean-square value
    of $\txir$ is unity. The plots are labeled with the corresponding
    eigenvalue $\tomega^{2}_{n}$; negative values for the $n=0$ and
    $n=1$ modes indicates that they are
    unstable.} \label{fig:eigfuncs}
\end{figure}

Figure~\ref{fig:eigfuncs} plots the dimensionless displacement
eigenfunctions, $\txir$, $\txit$, and $\txiz$, as a function of $\tr$
for the three lowest-order modes ($n = 0, 1, 2$). Each plot is labeled
at top-left by the corresponding eigenvalue $\tomega^{2}_{n}$. From
these eigenvalues, we see that the fundamental mode and first overtone
($n=1$) are both unstable, with $\tomega^{2} < 0$. From the
eigenfunctions, we see that the displacement in the azimuthal
direction is greater than that in the radial direction, $|\txit| >
|\txir|$, as predicted by \citet{Spruit_2002}. This makes sense
because displacement along the magnetic field lines on equipotential
gravitational surfaces does not need to do work against the
stratification.

\subsection{\protect Dependence on Axial Wavenumber} \label{ssec:instability_calcs_k}

\begin{figure}
  \centering
  \includegraphics{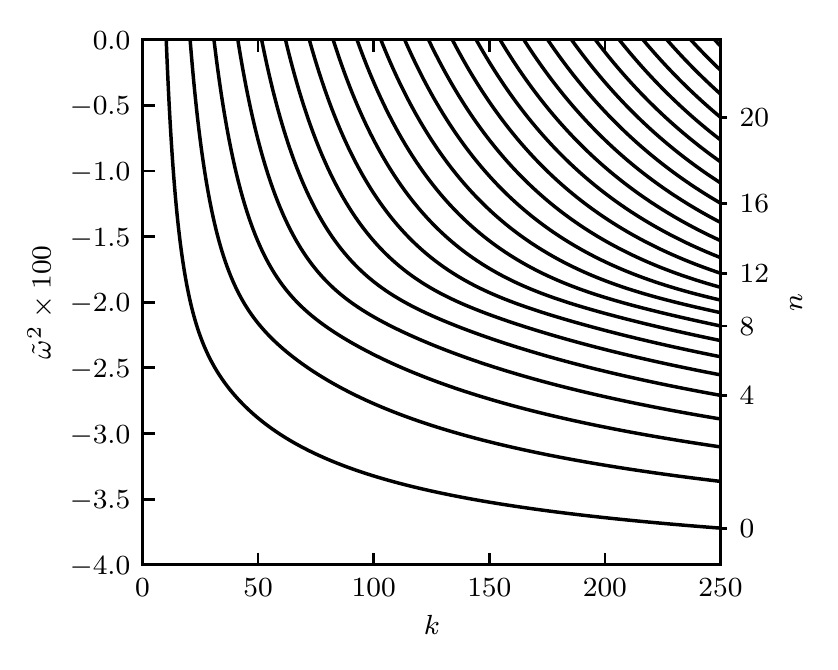}
  \caption{Eigenvalues $\tomega^{2}$ plotted as a
    function of axial wavenumber $k$, for an azimuthal wavenumber
    $m=1$ and an equilibrium model having $\beta=5$, $q=0.01$ and
    $\gamma=5/3$. Selected modes are labeled at right with their
    radial order $n$.} 
\label{fig:scan_k}
\end{figure}

We now explore how the mode eigenvalues depend on the axial wavenumber
$k$. With other parameters fixed at the values given in
Section~\ref{ssec:instability_calcs_modes}, Fig.~\ref{fig:scan_k}
plots the eigenvalues of unstable modes as a function of $k$. The plot
shows that for modes with a given radial order $n$ to become unstable, $k$ must exceed
some finite threshold $k_{n}$. Above this threshold, the eigenvalue
decreases monotonically with $k$, approaching an asymptotic limit as
$k \rightarrow \infty$. This is similar behavior to that found by
\citet{Pitts_Tayler_1985} for a uniform-density incompressible fluid
without gravity. In a non-ideal fluid with one or more forms of
diffusion (thermal, viscous, or resistive) diffusive damping is
expected to reduce the instability of modes at high $k$, leading to a
minimal $\tomega^{2} < 0$ (i.e., maximal growth rate) at large but
finite $k$.

\subsection{\protect Stability Boundaries} \label{ssec:instability_calcs_bounds}

Because the fundamental mode is the most unstable at any $k$, we can
use it as a proxy for the onset of the TI. Accordingly, we evaluate the minimal eigenvalue $\tomegamin^{2}$, over all modes, as
the limiting value of the fundamental-mode eigenvalue in the limit $k
\rightarrow \infty$:
\begin{equation} \label{eq:min_eigval}
  \tomegamin^{2} = \lim_{k\rightarrow\infty} \tomega_{0}^{2}.
\end{equation}
When $\tomegamin^{2} < 0$, the maximal exponential growth rate of the
Tayler instability is then given by
\begin{equation} \label{eq:growthmax}
  \growthmax = \frac{a |\tomegamin|}{r_{0}}.
\end{equation}
Because our numerical calculations are restricted to finite values of
$k$, we estimate the limit on the right-hand side of
equation~(\ref{eq:min_eigval}) by evaluating $\tomega_{0}^{2}$ at
$k=1,000$ and $k=2,000$, and then linearly extrapolating in inverse
wavenumber $k^{-1}$ to find the eigenvalue at $k^{-1} \rightarrow 0$.

\begin{figure}
  \centering
  \includegraphics{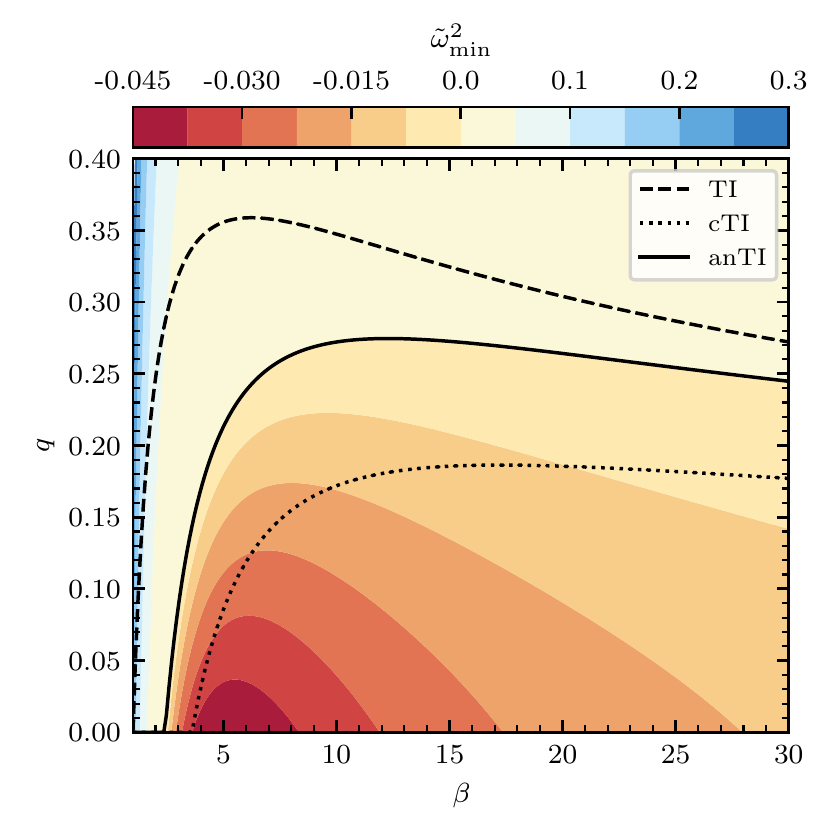}
  \caption{Contour map of the minimal eigenvalue $\tomegamin^{2}$,
    plotted across the $\beta$-$q$ plane for an azimuthal wavenumber
    $m=1$ and equilibrium models having $\gamma=5/3$. Regions where
    $\tomegamin^{2} < 0$ are unstable to the Tayler instability.  The
    three black lines show the stability boundaries for the fully
    compressible criteria (TI), the constrained anelastic criteria (cTI) and the
    LBR anelastic criteria (anTI). Only the LBR anelastic criteria correctly
    predict the $\tomegamin^{2}=0$ stability
    boundary.} \label{fig:map}
\end{figure}

We apply this approach to evaluate $\tomegamin^{2}$ for a grid of
equilibrium models spanning $1 \leq \beta \leq 30$ and $0.00 \leq q
\leq 0.40$. In all cases, we assume an azimuthal wavenumber $m=1$ and
$\gamma=5/3$. Fig.~\ref{fig:map} shows a contour map of the resulting
values. Plotted over the map are the stability boundaries
  predicted by the original stability criteria
  (equation~\ref{eq:coeff_tay_1d}), the cTI criteria
  (equation~\ref{eq:coeff_c_1d}) and the anTI criteria
  (equation~\ref{eq:coeff_lbr_1d}). The original criteria over-predict
  the extent of the instability seen in the numerical calculations,
  while the cTI criteria under-predict it. Only the anTI criteria
  correctly predict the stability boundary $\tomegamin^{2} = 0$,
  confirming that they are the correct choice for LBR anelastic MHD.

In his dynamo models, \citet{Spruit_1999,Spruit_2002} assumes that the
growth rate of the TI is of the order $\growth \approx \omegaalf$ in
the limit where the rotation angular frequency $\Omega$ is small compared to
the Alfv\'en frequency $\omegaalf$ defined by
\begin{equation} \label{eq:alf-freq}
\omegaalf^{2} = \frac{2 m^{2} a^2}{\beta r^2}.
\end{equation}
To compare this assumption against our calculations, we write the
ratio of growth rate to Alfv\'en frequency as
\begin{equation}
  \frac{\growth}{\omegaalf} = |\omega| \sqrt{\frac{\beta r^{2}}{2 m^{2} a^{2}}} \la |\tomegamin|\, \sqrt{\frac{\beta (\tra + \trb)^{2}}{8 m^{2}}},
\end{equation}
where the second (in)equality follows from setting $|\tomega| =
|\tomegamin|$, and evaluating the ratio at the midpoint $\tr = (\tra +
\trb)/2$ of the calculation domain. For modes with azimuthal order
$m=1$, we find this ratio has an average value $\lambda/\omegaalf
\approx 0.27$ over the unstable region plotted in Fig.~\ref{fig:map},
and a maximal value $\lambda/\omegaalf \approx 0.42$.  These are both
moderately smaller than the $\lambda/\omegaalf \approx 1$ assumed by
\citet{Spruit_1999,Spruit_2002}. Therefore, future studies involving
anelastic MHD simulations of the TI should recognize that the growth
rate can be smaller than $\omegaalf$, and adjust expectations
accordingly.

%% file: conclusion.tex
\section{Conclusion} \label{sec:conclusion}

The TI is of interest in the radiative regions of stars because,
with differential rotation, it may contribute to forming and
maintaining a magnetic field dynamo that could significantly affect the stars' structure and evolution.
However,
attempts to heuristically derive and numerically simulate the growth and saturation of the TI in stellar
models have led to indeterminate results.

The TI criteria were derived using fully compressible MHD, but
simulations of fluid dynamics in stellar interiors frequently use some
version of the anelastic or pseudo-incompressible approximations,
which suppress acoustic waves with much shorter periods than stellar
timescales. The goal of this paper was not to address the problem of whether the dynamo exists, but to narrow the gap between
fully-compressible linear theory and anelastic non-linear simulations of the TI.

We undertook this by modifying the classic MHD energy principle
\citep{Bernstein_etal_1958} according to LBR anelastic MHD, which ---
based on the work of \cite{Brown_etal_2012} --- we regard as the most
promising of several anelastic schemes. We derived a version of the
MHD energy principle that yields stability criteria
(equations~\ref{eq:coeff_lbr} and~\ref{eq:stab_lbr}) in
excellent agreement with solutions of the eigenvalue problem
calculated using the \gyre\ code. Our test configuration was a family of
cylindrically symmetric magnetohydrostatic equilibria with a toroidal
background magnetic field and gravity supplied by a line mass
(Section~\ref{ssec:instability_calcs_model}). Our results show that the instability still exists in LBR anelastic MHD, but in a more
restricted part of parameter space than the fully-compressible
case. This is because the energy principle is based on minimizing the
potential energy of the system, and anelasticity introduces a
constraint which precludes full minimization. However, we conclude
that the instability should manifest in anelastic LBR MHD simulations
if the models used are unstable under the anTI criteria.

We found that the amplitude of the displacement in the horizontal
direction is greater than the displacement in the radial direction, as
predicted by \cite{Spruit_2002}. We also found that the largest growth
rates calculated by \gyre\ are somewhat smaller than predicted for
slow rotators by \cite{Spruit_2002}.

We are limited in addressing discrepancies between our calculations
and heuristic predictions of the saturated state because our analysis
and numerical calculations are in the linear regime and lack rotation
or dissipative effects, both of which are key ingredients in the
proposed instability-driven dynamo \citep{Spruit_2002}. We are unable
to predict the non-linear growth rate and amplitude of the
instabilities without taking those physical effects into
consideration. That is beyond the scope of this work, but it is an
open question for future work.

Our family of cylindrical models can be implemented in anelastic MHD
simulations. Such simulations could verify the linear analysis and
calculations that we performed, and determine how non-linear effects
impact the growth rate and amplitude of the instability. Choosing
models that are unstable under the anTI criteria, and including
differential rotation, anelastic MHD simulations could more accurately
test the the Tayler-Spruit dynamo and its significance as a mechanism
for angular momentum transport in stellar evolution.

%% file: acknowledgements.tex
\section{Acknowledgements} \label{sec:acknowledgements}

Our work made possible through the collaborative effort of the
Supernova Progenitors, Internal Dynamics and Evolution Research
(SPIDER) network, supported via NASA TCAN program grant NNX14AB55G. We
also acknowledge support from NSF grants AST-1716436, PHY-1748958 and
ACI-1663696, and from the Wisconsin Alumni Research
Foundation. We thank Ryan Orvedahl and Benjamin Brown for their help
using Dedalus (http://dedalus-project.org) to verify the initial
\gyre\ calculations, and Erin Boettcher for her thorough review. EGZ
thanks the University of Chicago for hospitality during the completion
of this manuscript. We thank the referee for the useful comments that led to significant improvements in the paper.

%% file: appendix.tex
\appendix

\section{Numerical Technique} \label{sec:gyre-implement}

To calculate numerical solutions of the LBR anelastic
equations~(\ref{eq:lin_entropy}--\ref{eq:anelastic},\ref{eq:lbr_momentum}),
we first undertake a separation of variables in space and time, by
writing perturbed quantities in the form
\begin{gather}
  \begin{pmatrix}
    \xir  \\
    \xit  \\
    \xiz
  \end{pmatrix}
  =
  r_{0} \operatorname{Re} \left\{
    \begin{pmatrix}
          \txir  \\
      \ii \txit  \\
      \ii \txiz
    \end{pmatrix}
  \exp[ \ii ( m \phi + k z / r_{0} + \omega t ) ]
  \right\}
  \\
  \begin{pmatrix}
    \Br'  \\
    \Bt'  \\
    \Bz'
  \end{pmatrix}
  =
  \sqrt{\frac{8 \pi P_{0}}{\beta}} \operatorname{Re} \left\{
    \begin{pmatrix}
      \ii \tBr'  \\
          \tBt'  \\
          \tBz'
    \end{pmatrix}
  \exp[ \ii ( m \phi + k z / r_{0} + \omega t ) ]
  \right\}
  \\
  \begin{pmatrix}
    \Jr'  \\
    \Jt'  \\
    \Jz'
  \end{pmatrix}
  =
  \frac{c}{4\pi r_{0}} \sqrt{\frac{8 \pi P_{0}}{\beta}} \operatorname{Re} \left\{
    \begin{pmatrix}
      \ii \tJr'  \\
          \tJt'  \\
          \tJz'
    \end{pmatrix}
  \exp[ \ii ( m \phi + k z / r_{0} + \omega t ]
  \right\}.
  \\
  \Pi'  = \frac{P_{0}}{\rho_{0}} \operatorname{Re} \left\{ \tPi' \exp[ \ii ( m \phi + k z / r_{0} + \omega t ] \right\} \\
  S' = c_{P} \operatorname{Re} \left\{ \tS' \exp[ \ii ( m \phi + k z / r_{0} + \omega t ] \right\}
\end{gather}
In these expressions, all quantities with a tilde (\~{}) are
dimensionless real functions of $r$, to be determined numerically; the
integer $m$ is the azimuthal wavenumber introduced in
Section~(\ref{sec:instability_analysis_tay}), and the real number $k$
is the axial wavenumber. Here, we choose to work with
\begin{equation}
  \Pi' = \varpi' + \frac{B_{\phi} B_{\phi}'}{4\pi \rho_{0}}.
\end{equation}
rather than the reduced pressure $\varpi$, as this allows $\tJt'$ and
$\tJz'$ to be decoupled from the other dependent variables, reducing the
differential order of the system from four to two.

With these definitions, and assuming the equilibrium we introduce in
Section~\ref{ssec:instability_calcs_model}, we write the linearized equations
in the form
\begin{align}
  \label{eq:vw-diff}
  \frac{\diff \vv}{\diff \tr} &= \mA_{vv} \vv + \mA_{vw} \vw \\
  \label{eq:vw-alg}
    \boldsymbol{0}            &= \mA_{wv} \vv + \mA_{ww} \vw
\end{align}
where $\tr \equiv r/r_{0}$ is the independent variable, and the
vectors
\begin{equation}
  \vv = 
  \begin{pmatrix}
    \txir \\
    \tPi'
  \end{pmatrix},
  \qquad
  \vw =
  \begin{pmatrix}
    \txit  \\
    \txiz  \\
    \tBr'  \\
    \tBt'  \\
    \tBz'  \\
    \tJr'  \\
    \tS'
  \end{pmatrix}
\end{equation}
contain the dependent variables. The Jacobian matrices in
equations~(\ref{eq:vw-diff}) and~(\ref{eq:vw-alg}) are given by
\begin{equation}
  \mA_{vv} =
  \begin{pmatrix}
    \frac{\alpha - 1}{\tr} &
    0 \\
    \tomega^{2} &
    0
    \end{pmatrix},
\end{equation}
\begin{equation}
  \mA_{vw} =
  \begin{pmatrix}
    \frac{m}{\tr} &
    k &
    0 &
    0 &
    0 &
    0 &
    0 \\
    0 &
    0 &
    -\frac{2m}{\beta \tr} &
    \frac{2(\alpha - 2)}{\beta \tr} &
    0 &
    0 &
    \frac{q}{\tr}
  \end{pmatrix},
\end{equation}
\begin{equation}
  \mA_{wv} =
  \begin{pmatrix}
    0 &
    -\frac{m}{\tr} \\
    0 &
    -k \\
    \frac{m}{\tr} &
    0 \\
    \frac{2 - \alpha}{2 \tr} &
    0 \\
    0 &
    0 \\
    0 &
    0 \\
    \frac{\alpha(\gamma-1)}{\gamma \tr} &
    0
  \end{pmatrix},
\end{equation}
\begin{equation}
  \mA_{ww} =
  \begin{pmatrix}
    \tomega^{2} &
    0 &
    \frac{2 - \alpha}{\beta \tr} &
    \frac{2m}{\beta\tr} &
    0 &
    0 &
    0 \\
    0 &
    \tomega^{2} &
    0 &
    \frac{2k}{\beta} &
    0 &
    \frac{2}{\beta} &
    0 \\
    0 &
    0 &
    -1 &
    0 &
    0 &
    0 &
    0 \\
    -\frac{m}{\tr} &
    0 &
    0 &
    -1 &
    0 &
    0 &
    0 \\
    0 &
    -\frac{m}{\tr} &
    0 &
    0 &
    -1 &
    0 & 
    0 \\
    0 &
    0 &
    0 &
    -k &
    \frac{m}{\tr} &
    -1 &
    0 \\
    0 &
    0 &
    0 &
    0 &
    0 &
    0 &
    1
  \end{pmatrix},
\end{equation}
where we introduce the dimensionless frequency
\begin{equation} \label{eq:dim-freq}
  \tomega = \frac{r_{0}}{a} \omega.
\end{equation}

Eliminating $\vw$ between equations~(\ref{eq:vw-diff})
and~(\ref{eq:vw-alg}), we arrive at a system of differential equations
for $\vv$ alone:
\begin{equation} \label{eq:sys-v}
  \frac{\diff \vv}{\diff\tr} = \left( \mA_{vv} - \mA_{vw} \mA_{ww}^{-1} \mA_{wv} \right) \vv \equiv \mA \vv,
\end{equation}
where the second equality serves to define the overall Jacobian matrix
$\mA$. Although we do not write out an explicit expression for the
elements of $\mA$, we note that each contains a factor
\begin{equation}
F = \frac{1}{\beta \tr^{2} \tomega^{2} - 2 m^{2}} = \frac{1}{\beta \tr^{2} (\tomega^{2} - \tomegaalf^{2})},
\end{equation}
where
\begin{equation} \label{eq:dim-alf-freq}
\tomegaalf = \frac{r_{0}}{a} \omegaalf
\end{equation}
is the dimensionless equivalent of the Alfv\'en frequency defined in
equation~(\ref{eq:alf-freq}). The factor $F$ diverges if $\tomega
= \tomegaalf$, indicating a local resonance with the Alfv\'en wave. In
the present context, such behavior is not a problem because we are
interested in finding unstable modes for which $\tomega^{2} < 0$, and
therefore the resonance never arises.

Together with the boundary conditions
\begin{equation} \label{eq:bound-v}
  v_{1} = \txir = 0
\end{equation}
on the inner ($\tr = \tra$) and outer ($\tr = \trb$) boundaries of the
calculation domain (in accordance with equation~\ref{eq:boundary}),
the system of equations~(\ref{eq:sys-v}) is a linear two-point
boundary eigenvalue problem (BVEP), with $\tomega^{2}$ serving as the
eigenvalue.  To solve the BVEP numerically we use the \gyre\
code \citep{Townsend_Teitler_2013}. Although \gyre\ is designed to
address stellar pulsation problems, it is built on a robust
multiple-shooting scheme which can in principle be applied to any
BVEP. Accordingly, we modify \gyre\ to implement the differential
equations and boundary conditions given here. The modified code takes
as inputs parameters specifying the equilibrium model
($\beta,\gamma,q$), the wavenumbers ($m,k$), and the calculation
domain ($\tra, \trb$, and the number of points $N$ used to discretize
the differential equations). As outputs, it calculates the eigenvalues
$\tomega^{2}$ of the discrete modal solutions, and the corresponding
eigenfunctions given by the components of $\vv$ and $\vw$.